\documentclass{jfm_clean}

\usepackage{graphicx}
\usepackage{newtxtext}
\usepackage{newtxmath}
\usepackage{natbib}
\usepackage{hyperref}
\hypersetup{
	colorlinks = true,
	urlcolor   = blue,
	citecolor  = black,
}

\newcommand{\RomanNumeralCaps}[1]
\linenumbers

\title{Water wave interactions with a horizontal submerged elastic plate}
\author{Gatien Polly\aff{1},
	Alexis Mérigaud\aff{1,2},
	Benjamin Thiria\aff{1}
	\and Ramiro Godoy-Diana \aff{1}	\corresp{\email{ramiro@pmmh.espci.fr}}}
\affiliation{\aff{1} Laboratoire de Physique et Mécanique des Milieux Hétérogènes (PMMH), CNRS UMR 7636,
	ESPCI~Paris--Université PSL, Sorbonne Université, Université Paris Cité, 75005 Paris, France
	\aff{2} IFP Energies Nouvelles, 92852 Rueil-Malmaison, France}

\begin{document}
	\maketitle
	
	\begin{abstract}
	This article explores how a submerged elastic plate, clamped at one edge, interacts with water waves. Submerged elastic plates have been considered as potentially effective design elements in the development of wave energy harvesters but their behavior in a wave field remains largely unexplored, especially experimentally. Positioned at a fixed depth in a wave tank, the flexible plate demonstrates significant wave reflection capabilities, a characteristic absent in rigid plates of identical dimensions. The experiments thus reveal that plate motion is crucial for wave reflection. Sufficiently steep waves are shown to induce a change in the mean position of the plate, with the trailing edge reaching the free surface in some cases. This configuration change is found to be particularly efficient to break water waves. These findings contribute to understanding the potential of elastic plates for wave energy harvesting and wave attenuation scenarios.
	\end{abstract}

\thispagestyle{empty}	
	
%	\begin{keywords}
%		Authors should not enter keywords on the manuscript, as these must be chosen by the author during the online submission process and will then be added during the typesetting process (see \href{https://www.cambridge.org/core/journals/journal-of-fluid-mechanics/information/list-of-keywords}{Keyword PDF} for the full list).  Other classifications will be added at the same time.
%	\end{keywords}
%	
%	{\bf MSC Codes }  {\it(Optional)} Please enter your MSC Codes here
	\section{Introduction}

	The potential of wave energy as a renewable electricity source has been assessed to exceed the global demand for electricity \citep{Sheng2019, Mork2010}. However, efficient designs still remain to be developed at an industrial level. Horizontally submerged flexible plates could be part of those designs due to their improved survivability \citep{Collins2021}.
	Until now, flexible submerged elastic plates have mostly been studied numerically and theoretically in a configuration where both edges are clamped, with the objective of harnessing wave energy or protecting the coastline. \cite{Cho1998} and \cite{Cho2000} considered the plate as a potential wave barrier. From the on-site observations that a muddy seafloor can effectively dampen water waves, \cite{Alam2012} proposed a design to harvest wave energy by mimicking this phenomenon. To that aim,  a flexible plate is used as a wave energy harvester. The flexible plate is hinged at the seafloor and oscillates due to the wave forcing, similarly to the mud in the ocean floor. Those oscillations are used to produce electricity by means of a Power-Take-Off (PTO) system. The Wave-Carpet design has been investigated numerically and experimentally by \cite{Desmars2018}, \cite{Asaeian2020} and \cite{Lehmann2013}. \cite{Renzi2016} studied numerically a variation of the Wave-Carpet considering the use of a piezoelectric material to produce electricity, while \cite{Achour2020} added the effect of a current in addition to the waves. Finally, submerged flexible plates clamped at both edges have been studied in more complex configurations, such as close to a wall as potential wave barriers \citep{Guo2020, Gayathri2020, Mohapatra2020}, superimposed with other plates \citep{Behera2021, Mohapatra2014, Mohapatra2019} or in series \citep{Mirza20241, Mirza20242}.

	Contrary to the two edge clamping, clamping the plate at only one edge has barely been studied. \cite{Shoele2023} investigated numerically the case of a submerged elastic plate as a hybrid current/wave energy converter but the impact of waves alone has not been investigated. The aim of this paper is to characterize experimentally the interaction of a flexible submerged elastic plate, clamped at only one edge, with water waves.
	A table top experiment has been developed to study the wave-plate interaction, which allows the use of vision-based data extraction techniques. Schlieren imaging is utilized to measure the free surface height and a side view of the tank to track the plate deformation.
	The ability of the plate to reflect, transmit or dissipate water waves is computed measuring the reflection and transmission coefficients. It is observed that the flexible plate can reflect water-waves over a certain range of waves frequencies. It exposes potential applications of such a system for coastal protection and, more generally, offers a deeper fundamental understanding of the role played by flexibility in the wave-structure interaction applications. The comparison of wave reflection and transmission by a submerged flexible plate and a submerged rigid plate of the same dimensions shows that the plate motion is responsible for wave reflection. Finally, when increasing the wave amplitude, the flexible plate can act like a perfect wave absorber by changing its mean position. The flexible plate acts as a reconfigurable water wave absorber which could be able to totally dampen the waves in high sea conditions.
	
	\section{Experimental methods}
	\begin{figure}
		\centerline{\includegraphics[width=\linewidth]{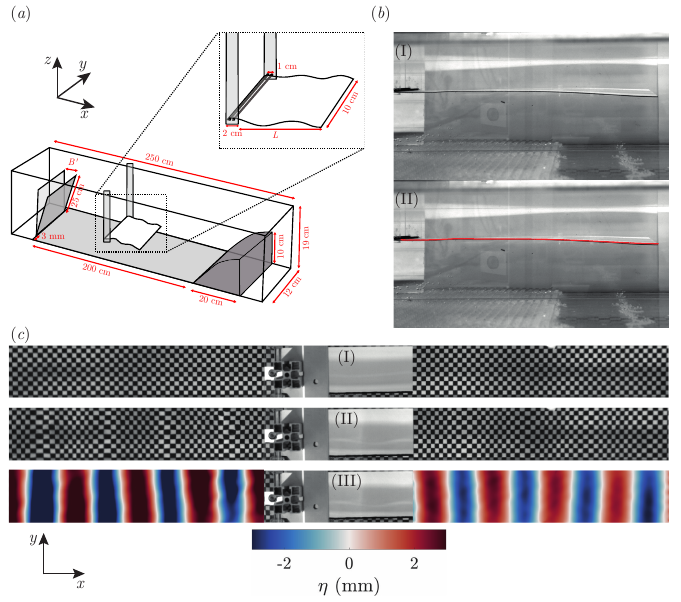}}% Images in 100% size
		\captionsetup{width=1\linewidth}
		\caption{(\textit{a}) Schematic diagram of the experimental set-up. The plate length $L$ is 28~cm. (\textit{b}) Side view of the tank with the plate: Picture (I) is a raw image while picture (II) illustrates the detected edge of the plate (red line). (\textit{c}) Tank top view: Picture (I) is taken with an undisturbed free surface. For picture (II) 3.5~Hz/13~cm waves of 3~mm amplitude are generated and the checkerboard appears deformed. The bottom picture illustrates the free surface height reconstruction using Schlieren imaging.}
		\label{Fig: Method}
	\end{figure}
	\subsection{Experimental setup}
	The experimental setup is represented in Figure~\ref{Fig: Method}~(\textit{a}), in a reference frame with $x$ the direction of propagation of the waves, $y$ the transverse direction, and $z$ the vertical coordinate, respectively. The total length of the tank is 2.5~m, where the measurement region is 2~m long at the center of the tank. The tank width is restricted to a 12~cm channel to prevent transverse modes in the investigated range of frequencies \citep{Ursell1952}. Water depth is fixed at 10~cm. Waves are produced using a flap type wave-maker driven by a linear motor (DM01-23x80F-HP-R-60\textunderscore MS11 from Linmot\textregistered). In practice, the waves produced in this setup have an amplitude ranging from 0.5 to 5~mm and frequencies ranging from 1.5~Hz to 4~Hz. The wave angular frequency, $\omega$, is related to the wavenumber, $k$, through the dispersion relation of gravity waves:
	\begin{equation}
		\omega^2=gk\tanh(kh),
	\end{equation}  
	where $g$ is the acceleration of gravity and $h$ the water depth. The studied wavelengths, $\lambda$, range from 10~cm to 55~cm, so that $\lambda/h$ ranges from 1 to 5.5, corresponding to deep to intermediate water depths.\\
	The plate is positioned halfway through the tank length at $x=0$ m. At the end of the tank, a parabolic beach is placed. It is used to limit reflection on the tank wall \citep{Ouellet1986}. The beach is pierced with 5~mm circular holes to enhance viscous damping. 
	Data is extracted from this setup by filming simultaneously from the top and the side of the tank, giving access to the free surface deformation and the plate motion.

	\subsection{Water wave height measurements}

	The free surface deformation is computed using Schlieren Imaging \citep{Moisy2009,Wildeman2018}. To do so, a checkerboard pattern is placed at the bottom of the tank. When waves propagates, the image of the checkerboard obtained by filming with the top camera appears to be deformed, as illustrated in Figure~\ref{Fig: Method}~(\textit{c}). Both Figures~\ref{Fig: Method}~(\textit{c})~(I)~and~(II) show an image taken by the top camera with the plate placed in the middle of the tank. In (I), no waves are propagating, whereas in (II) 3.5~Hz/13~cm waves of 2.5~mm amplitude are propagating in the tank. The comparison between those two images is used to reconstruct the free surface height in all the tank by using the open-source code provided by \cite{Wildeman2018}. Figure~\textit{(c)}~(III) illustrates free surface reconstruction using the top and middle images.\\
	The tank can be divided into two different regions: the up-wave region, located between the wave-maker and the plate, where the incoming wave, produced by the wave-maker, and the wave reflected by the plate propagate; and the down-wave region, located between the plate and the beach, where waves transmitted by the plate propagate, as well as a small component of waves reflected from the absorption beach.
	After filtering at the wave-maker angular frequency, $\omega$, the free surface elevation can be described as the sum of two waves propagating forward and backward in each region:
		\begin{equation}
		\left\{
		\begin{array}{lll}
			\eta^{\mathrm{uw}}(x,t)=\mathrm{Re}(\hat\eta^{\mathrm{uw}}_+(x)e^{-i\omega t}+\hat\eta^{\mathrm{uw}}_-(x)e^{-i\omega t})~~\mathrm{with}~~
			\hat\eta^{\mathrm{uw}}_\pm(x)=\hat A_\pm^{\mathrm{uw}}e^{\pm(ik-\nu)x }\\
			\\
			\eta^{\mathrm{dw}}(x,t)=\mathrm{Re}(\hat\eta^{\mathrm{dw}}_+(x)e^{-i\omega t}+\hat\eta^{\mathrm{dw}}_-(x)e^{-i\omega t})~~\mathrm{with}~~
			\hat\eta^{\mathrm{dw}}_\pm(x)=\hat A_\pm^{\mathrm{dw}}e^{\pm(ik-\nu)x },\\
		\end{array}
		\right.\label{Eq: expression FS}
	\end{equation}
	where $\eta$ is the free surface height and the superscripts uw and dw stand for up-wave and down-wave, respectively. Quantities associated with waves propagating forward and downward are indicated using the subscripts + and -, respectively, and complex numbers are denoted using the symbol $~\hat{ }$. $\hat\eta$ is the complex spatial part of the free surface height, it depends on $k$ the wavenumber, $\hat A$ the complex wave amplitudes, and $\nu$ a damping coefficient that models wave dissipation along the tank.
	
	The energy reflection and transmission coefficients, $K_r$ and $K_t$, for a given wave amplitude and frequency are defined as:
	\begin{equation}
		K_r=\left(\frac{|\hat A_-^{\mathrm{uw}}|}{|\hat A_+^{\mathrm{uw}}|}\right)^2~\mathrm{and}~K_t=\left(\frac{|\hat A_+^{\mathrm{dw}}|}{|\hat A_+^{\mathrm{uw}}|}\right)^2.
	\end{equation}
	
	To determine the reflection and transmission coefficients, the amplitude of the waves propagating forward and backward have to be measured. 
	The image deformation is related to the free surface height by \citep{Moisy2009}:
	\begin{equation}
		\nabla\eta=-\frac{\mathbf{u}}{h^*},\label{Eq: lien FS def}
	\end{equation}
	where $h^*$ is a real-valued parameter that depends on water depth and setup configuration and $\mathbf{u}$ is the deformation field. The vector $\mathbf{u}$ has two components, $u_x$ and $u_y$, corresponding to the image deformation along $x$ and $y$. Generally, $\eta$ is obtained by numerically inverting the gradient. However, as the free surface height is non zero on the edges of the image, the determination of the integration constant leads to numerical errors. In this work, gradient inversion is avoided, by using the monochromatic nature of the waves studied. The elevation gradient, $\nabla\hat\eta$, is reduced to a simple partial derivative along $x$. 
	Equation~\ref{Eq: expression FS}, leads to:
	\begin{equation}
		\nabla\hat\eta_+=(ik-\nu)\hat\eta_+~~\mathrm{and}~~\nabla\hat\eta_-=-(ik-\nu)\hat\eta_-
	.\end{equation}
The Fourier transform of $u_x$ at the wave-maker frequency, $\hat u_x(x)$, can also be decomposed in its forward/backward components as:
\begin{equation}
	\hat u_x(x)=\hat u_{x+}e^{(ik-\nu )x}+\hat u_{x-}e^{(-ik+\nu )x},
\end{equation}
with:
\begin{equation}
	\hat u_{x+}=-(ik-\nu)h^*\hat\eta_+~~\mathrm{and}~~\hat u_{x-}=(ik-\nu)h^*\hat\eta_-.
\end{equation}
Therefore, for monochromatic waves, $K_r$ and $K_t$ can be computed using only the image deformation:
	\begin{equation}
		K_r=\left(\frac{|\hat u_{x-}^{\mathrm{uw}}|}{|\hat u_{x+}^{\mathrm{uw}}|}\right)^2~\mathrm{and}~K_t=\left(\frac{|\hat u_{x+}^{\mathrm{dw}}|}{|\hat u_{x+}^{\mathrm{uw}}|}\right)^2.
	\end{equation}
	The four deformation amplitudes are obtained by choosing an arbitrary set of positions $x_i$ where $\hat u_x$ is evaluated. The different amplitudes and the damping coefficient are determined by solving the system: 
	\begin{equation}
		\left\{
		\begin{array}{lll}
			\hat u_x(x_1)=\hat u_{x+}\exp(ikx_1-\nu x_1)+\hat u_{x-}\exp(-ikx_1+\nu x_1)\\
			\vdots\\
			\hat u_x(x_{\mathrm{max}})=\hat u_{x+}\exp(ikx_{\mathrm{max}}-\nu x_{\mathrm{max}})+\hat u_{x-}\exp(-ikx_{\mathrm{max}}+\nu x_{\mathrm{max}}),
		\end{array}
		\right.\label{Eq: system NLS}
	\end{equation}
	using non-linear least-squares method.
	
	Top and side views movies are taken simultaneously by following the same protocol. First, a picture of the undisturbed tank is taken, giving a reference image for the top and side view. Then, waves are sent for one minute and fifteen seconds. The first minute is used to reach a steady state and data is acquired during the last 15~s.

	\subsection{Plate characteristics and clamping system}
	
	The plate is placed submerged as pictured in Figure~\ref{Fig: Method}, at a fixed depth of 3~cm. The leading edge of the plate is maintained submerged using a clamping system composed of two carbon rods of 2~mm diameter that are glued to the edge of the plate. The rods are attached to the support poles on the sides of the tank, ensuring that the leading edge remains still. The plate material choice is particularly crucial. Indeed, since the plate trailing edge remains free, a small density difference with water will induce a deviation from the horizontal at rest. In practice, such deviation can be mitigated by using a relatively rigid material. The plate itself is made by cutting polypropylene sheets of density 1035~kg.m$^{-3}$ and stiffness 1.7.10$^{-3}$~N.m$^{2}$. 
	
	Figure~\ref{Fig: Method}~(\textit{b})~(I) is taken from the tank side view and shows the plate when no waves are present. The plate edge is colored in black to allow its tracking as illustrated in the picture (II). As can be appreciated in the Figure, the plate position is close to horizontal.
	The plate length, $L$, is fixed at 28~cm, which enables to test ratios $L/\lambda$ ranging from 0.45 to 2.7. The plate resonance frequencies have been measured forcing the plate at a 3~cm depth. For forcing frequencies between 0.5~Hz and 4~Hz, two resonance frequencies of the system are observed, corresponding to the second and third plate mode, at respectively $f_2=0.6$~Hz and $f_3=1.96$~Hz.

		\begin{figure}
		\centerline{\includegraphics[width=1\linewidth]{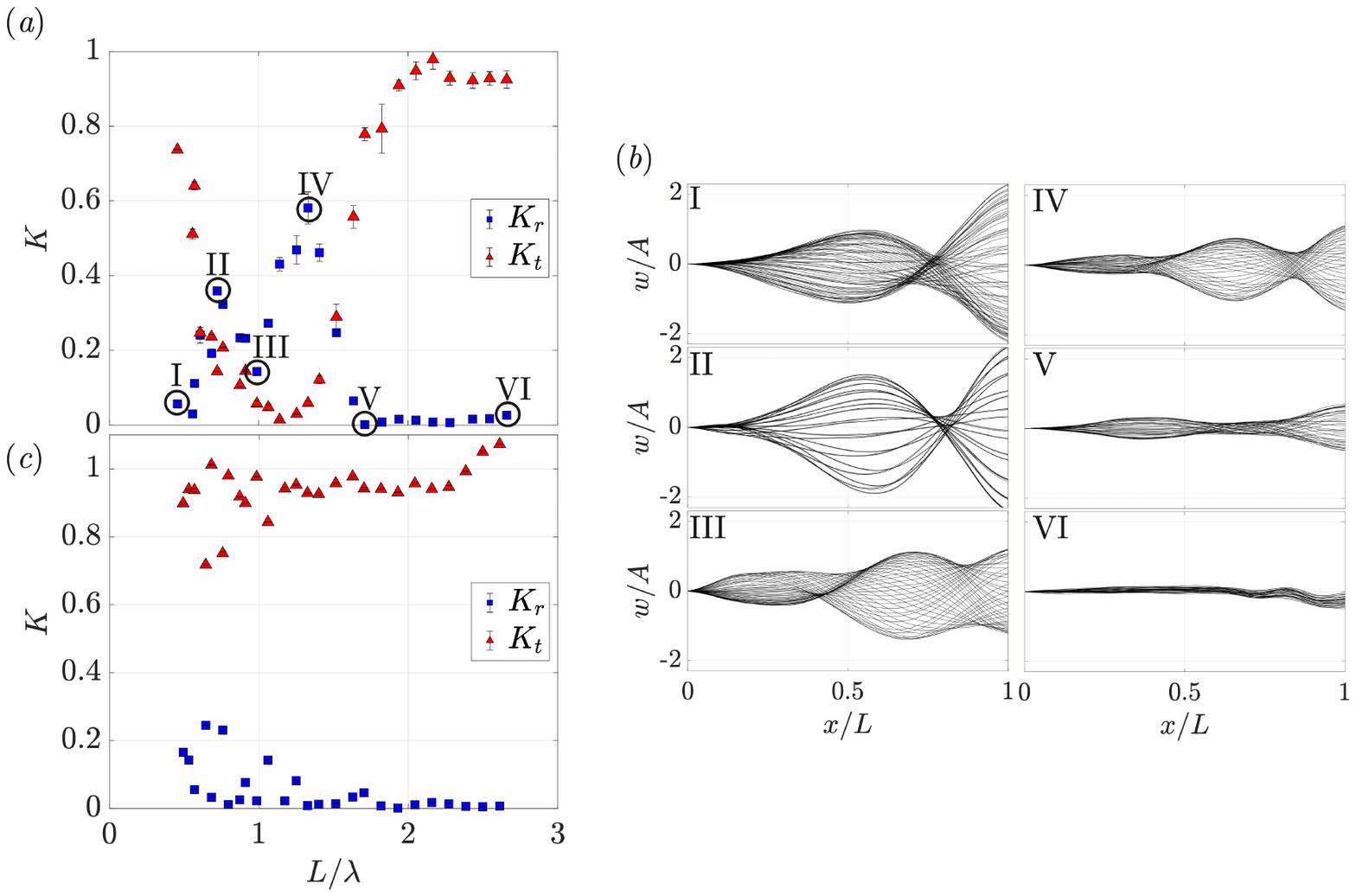}}% Images in 100% size
		\captionsetup{width=1\linewidth}
		\caption{(\textit{a}) Average over three experiments of the reflection and transmission coefficients, $K_r$ (blue squares) and $K_t$ (red triangles) for $L/\lambda$ from 0.45 to 2.7. A significant reflection is observed for $L/\lambda$ between 0.6 and 1.6 while for other values the plate mostly transmits waves. (\textit{b}) Plate deflection $w$ normalized by the incoming wave amplitude, $A$, for six values of $L/\lambda$ indicated by roman numbers from I to VI in Figure (\textit{a}). Across the different values of $L/\lambda$, the plate amplitude of motion decreases and its deformation mode changes. (\textit{c}) Reflection (blue squares) and transmission (red triangles) coefficients for a rigid plate of same dimension as the flexible one. No reflection is observed underlining the fact that the wave reflection is induced by the plate motion.}
		\label{Fig: Results_lowA}
	\end{figure}
			\begin{figure}
		\centerline{\includegraphics[scale=0.7]{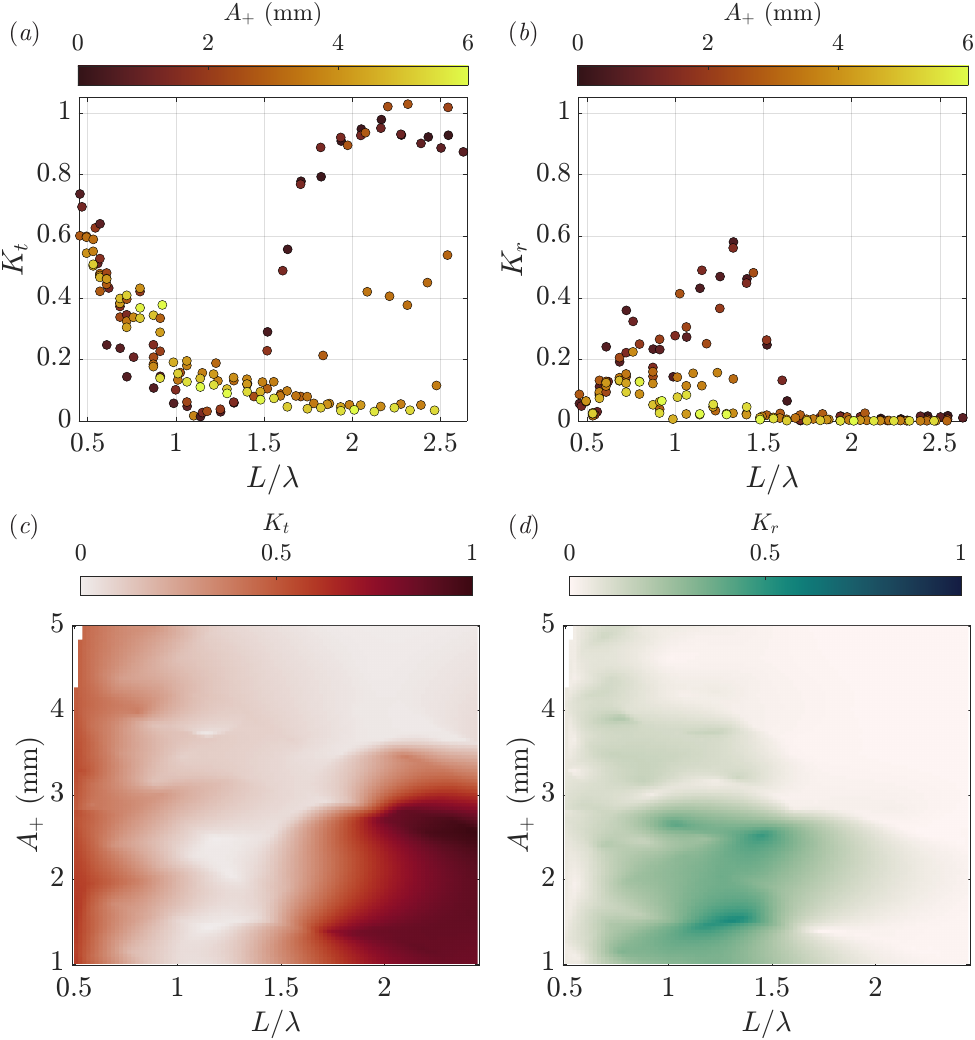}}% Images in 100% size
		\captionsetup{width=1\linewidth}
		\caption{(\textit{a}) Transmission coefficient, $K_t$, as a function of $L/\lambda$ for six different wave amplitudes. The wave amplitude is indicated with different colors, from black to yellow, yellow corresponding to higher wave amplitudes. At low $L/\lambda$, the wave amplitude has few impact on $K_t$, while for high amplitude waves, the transmission coefficient drops to zero. (\textit{b}) Reflection coefficient, $K_r$, as a function of $L/\lambda$ for six different wave amplitudes, using the same color code as for the transmission coefficient. The increase in wave amplitude leads to a decrease in transmission. (\textit{c})-(\textit{d}) $A-L/\lambda$ diagrams showing the transmission (\textit{c}) and reflection coefficients (\textit{d}) interpolated using experimental data to cover the whole domain. Bright colors correspond to the coefficient being equal to 0 and dark colors to the coefficient equal to 1. For the steepest waves, no reflection and no transmission are observed, meaning that the plate dissipates all the incoming wave energy.}
		\label{Fig: RT_bigA}
	\end{figure}
	\section{Results}

	\subsection{Low wave amplitude forcing \label{Sec: RLowA}}

	Figure~\ref{Fig: Results_lowA}~(\textit{a}) shows the average, over three experiments, of the reflection and transmission coefficients, $K_r$ and $K_t$, measured as a function of the ratio $L/\lambda$. These experiments were conducted with a water wave amplitude of approximately 0.5~mm, which corresponds to the lowest wave amplitude that can be produced and effectively analyzed. At lower values of $L/\lambda$ waves are mostly transmitted by the plate with the reflection coefficient, $K_r$, being almost equal to 0 for point (I). Increasing $L/\lambda$ leads $K_t$ to drop to 0 and $K_r$ to rise to 60~\% of the incoming wave energy (IV). For $L/\lambda$ larger than 1.6, $K_r$ drops again to 0 and the plate mostly transmits waves. It is observed that for $L/\lambda<1.6$, $K_r+K_t<1$, implying that a significant part of the incoming wave energy is dissipated.
	Figure~\ref{Fig: Results_lowA}~(\textit{b}) shows the plate deflection over one wave period, normalized by the incoming wave amplitude for different values of $L/\lambda$. When increasing $L/\lambda$, the plate amplitude of motion decreases. Also, when $L/\lambda$ increases, the plate changes its deformation mode. It is illustrated by Figure~\ref{Fig: Results_lowA}~(\textit{b})~(I) and (II) that have one node, while (III) and (IV) show two nodes. As illustrated by cases (V) and (VI), the values of $L/\lambda$, where the plate fully transmits water wave, is associated with almost no plate motion. 
	
	The reflection and transmission are also examined using a rigid plate made of aluminum having the same dimensions as the flexible plate. The corresponding $K_r$ and $K_t$ are presented in Figure~\ref{Fig: Results_lowA}~(\textit{c}) as a function of $L/\lambda$. No reflection pattern is observed using a rigid plate, highlighting the fact that the plate motion is necessary to reflect water waves in this case.
	\\
	From a frequency perspective, the reflection zone is centered on 2.5~Hz (\textit{i.e.} $f/f_2$=4 and $f/f_3$=1.25), covering frequencies between 1.8 and 3~Hz (\textit{i.e.} $f/f_2\in$[3-5] and $f/f_3\in$[0.9-1.5]). 
	\begin{figure}
	\centerline{\includegraphics[scale=0.9]{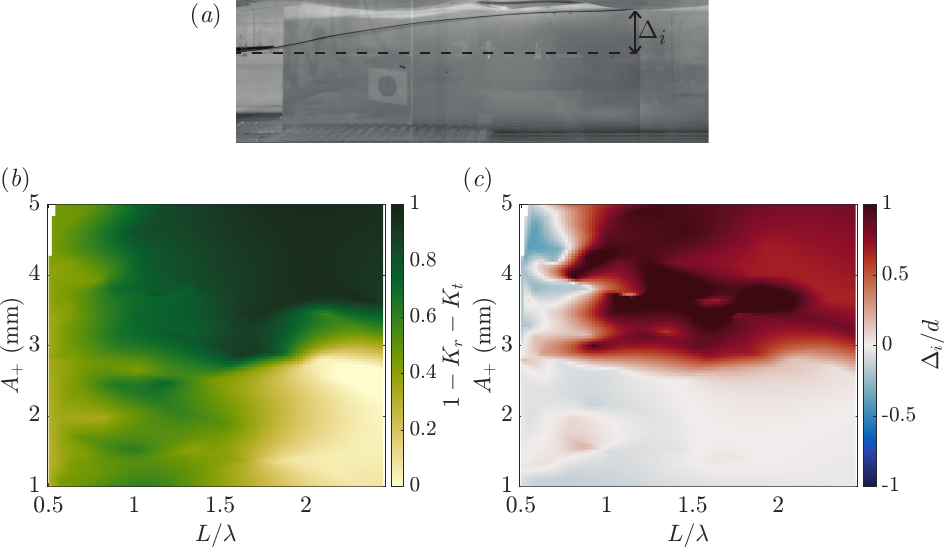}}% Images in 100% size
	\captionsetup{width=1\linewidth}
	\caption{(\textit{a}) Side view of the tank while sending waves of 6~mm amplitude with $L/\lambda$=0.99. The plate tip is displaced from its initial position of $\Delta_i$ and reaches the free surface. (\textit{b}) $A-L/\lambda$ diagram showing $1-K_r-K_t$, meaning the part of the incoming wave energy that is dissipated by the interaction with the plate. Dissipation is maximum for the steepest waves.  (\textit{c}) $A_+-L/\lambda$ diagram showing the plate tip mean displacement $\Delta_i$ normalized by plate depth, $d$. For the steepest waves, $\Delta_i\approx d$, meaning that the plate tip reaches the free surface.}
	\label{Fig: plate position}
\end{figure}
	\subsection{Large wave amplitude forcing}

	Figures~\ref{Fig: RT_bigA} (\textit{a}) and (\textit{b}) illustrate the influence of the incoming wave amplitude, $A_+$, on the transmission and reflection coefficients, respectively. At lower $L/\lambda$, the wave amplitude has no influence on wave transmission by the plate. However, for larger $L/\lambda$, a drop in transmission is observed when $A_+$ increases. In terms of reflection, the increase in the incoming wave amplitude leads to an attenuation of the reflection peak.\\ By interpolating all the experimental data points, we produce both the $K_r$ and $K_t$ maps in a wave-amplitude-$L/\lambda$ space shown in
	figures~\ref{Fig: RT_bigA} (\textit{c}) and (\textit{d}). The lower amplitude case corresponds to data presented in the Section~\ref{Sec: RLowA} with significant reflection for $L/\lambda$ between 0.6 and 1.6. When increasing the wave amplitude, the reflection and transmission patterns are drastically modified. For $A_+>3$~mm waves are neither reflected nor transmitted for $L/\lambda>0.6$, meaning that the incoming wave energy is totally dissipated when interacting with the plate. This dissipation can be attributed to a change in the plate mean position for the $(A_+-L/\lambda)$ couples that are illustrated in Figure~\ref{Fig: RT_bigA}. Figure~\ref{Fig: plate position}~(\textit{a}) shows a picture taken from the side of the tank when conducting experiments for $A_+=6$~mm and $L/\lambda=0.99$. The plate tip reaches the  free surface under the effect of the waves. Similarly to a beach, this configuration is found to be particularly efficient to break water waves. The change in plate tip mean position, $\Delta_i$, can be measured for all the experiments. Its values are presented in Figure~\ref{Fig: plate position}~(\textit{b}). The region where both $K_r$ and $K_t$ are equal to 0 corresponds to $\Delta_i/d\approx$1, \textit{i.e.} experiments for which the plate tip reaches the free surface.
	Figure~\ref{Fig: plate position}~(\textit{c}) illustrates the correlation between the plate tip position and the part of the incoming wave energy that is dissipated, $1-K_r-K_t$. The region where the plate tip reaches the free surface corresponds to the region where all the energy of the incoming wave is dissipated because of the interaction with the plate \textit{i.e.} $1-K_r-K_t=1$. For higher $L/\lambda$ and lower amplitudes, no energy is dissipated, which is associated with regimes where the plate does not move.

	\section{Discussion and conclusion}
	
	The present study characterizes the interaction between a flexible plate, clamped at one edge, with water waves. It highlights experimentally that in such a configuration, the flexible plate can reflect wave energy. The comparison with a rigid plate enlightens the key role played by the plate's flexibility for potential coastal protection applications. The role of flexibility in structures intended for coastal protection has been largely discussed and is known to enhance wave dissipation compared to rigid structures \citep{Kumar2007}. However, the influence of flexibility on wave reflection remains generally unclear. For instance, \cite{Stamos2001} reported higher reflection coefficients for rigid submerged breakwaters. The present work highlights the opposite behavior for a submerged elastic plate.
	
	For this system, an uncommon physical process  enhances wave dissipation. Owing to its ability to change its mean position when forced by the waves, the plate is passively placed in a configuration extremely efficient to break water waves. This configuration change, where the plate tip reaches the free surface, intervenes for the steepest water waves. The minimal steepness at which a significant change in position occurs is $A_+/\lambda\approx 10^{-2}$, for $A_+=3$~mm and $L/\lambda\approx1$, which corresponds to slightly nonlinear waves \citep{LeMehaute1976}. Strong nonlinear behaviors can therefore be observed for waves that are far from the breaking limit. Several phenomena could be responsible for the plate changing its mean position. First, if a current takes place between the plate and the free surface, it could cause a depression similar to a Venturi effect. In such a configuration, Stokes drift could arise from wave steepness \citep{Stokes1847, Van2018}. In addition, waves above a rigid plate can lead to the creation of a mean flow \citep{Carmigniani2017}. The flow at the edge of the plate could explain the observations through its associated upward force. In the case of a rigid plate in a wave field, vortex creation has been reported by \cite{Boulier1994}, \cite{Poupardin2012} and \cite{Pinon2017}. They observed that the vortex street has a given direction, slightly upward or downward. A similar phenomenon could  exert a force upward on the plate tip.
	
	Finally, this study opens questions regarding the origin of wave reflection by the plate. It appears that the reflection observed here cannot be completely predicted using a simple unique physical parameter. For instance, neither the plate length nor the plate resonance frequencies account for the reflection. Simulations, similar to the work of \cite{Shoele2023} or \cite{Renzi2016}, could be efficient tools to examine this phenomenon through a wider exploration of the parameter space.

\bibliographystyle{jfm}
%\bibliography{biblio2}

\begin{thebibliography}{33}
\expandafter\ifx\csname natexlab\endcsname\relax\def\natexlab#1{#1}\fi
\def\au#1{#1} \def\ed#1{#1} \def\yr#1{#1}\def\at#1{#1}\def\jt#1{\textit{#1}}
  \def\bt#1{#1}\def\bvol#1{\textbf{#1}} \def\vol#1{#1} \def\pg#1{#1}
  \def\publ#1{#1}\def\arxiv#1{#1}\def\org#1{#1}\def\st#1{\textit{#1}}

\bibitem[Achour {\em et~al.\/}(2020)Achour, Mougel, Lo~Jacono \&
  Fabre]{Achour2020}
{\sc \au{Achour, N.}, \au{Mougel, J.}, \au{Lo~Jacono, D.} \& \au{Fabre, D.}}
  \yr{2020}  \at{Etude théorique de l'effet d’un faible courant sur les
  interactions houle/membrane flexible: application à la recuperation
  d’énergie.}  \jt{Revue Paralia}  \pg{pp. 495--504.}

\bibitem[Alam(2012)]{Alam2012}
{\sc \au{Alam, M.-R.}} \yr{2012}  \at{Nonlinear analysis of an actuated
  seafloor-mounted carpet for a high-performance wave energy extraction}.
  \jt{Proc. R. Soc. A: Math. Phys. Eng.}  \bvol{468}~(2146),  \pg{3153--3171}.

\bibitem[Asaeian {\em et~al.\/}(2020)Asaeian, Abedi, Jafari-Talookolaei \&
  Attar]{Asaeian2020}
{\sc \au{Asaeian, A.}, \au{Abedi, M.}, \au{Jafari-Talookolaei, R.-A.} \&
  \au{Attar, M.}} \yr{2020}  \at{Wave propagation through a submerged
  horizontal plate at the bottom of a water channel}.  \jt{Comput. Eng. Phys.
  Model.}  \bvol{3}~(4),  \pg{1--19}.

\bibitem[Behera(2021)]{Behera2021}
{\sc \au{Behera, H.}} \yr{2021}  \at{Oblique wave scattering by a system of
  semi-infinite floating and submerged elastic plates}.  \jt{Differ. Equ. Dyn.
  Syst.}  \bvol{29}~(1),  \pg{157--173}.

\bibitem[Boulier \& Belorgey(1994)]{Boulier1994}
{\sc \au{Boulier, B.} \& \au{Belorgey, M.}} \yr{1994}  \at{Ecoulement
  tourbillonnaire et zone d’affouillement g{\'e}n{\'e}r{\'e}s par la houle en
  pr{\'e}sence d’une plaque immerg{\'e}e}.  \jt{3{\`e}mes JNGCGC. S{\`e}te,
  France}  \pg{pp. 39--45}.

\bibitem[Van~den Bremer \& Breivik(2018)]{Van2018}
{\sc \au{Van~den Bremer, T.~S.} \& \au{Breivik, {\O}.}} \yr{2018}  \at{Stokes
  drift}.  \jt{Philos. trans., Math. phys. eng. sci.}  \bvol{376}~(2111),
  \pg{20170104}.

\bibitem[Carmigniani {\em et~al.\/}(2017)Carmigniani, Benoit, Violeau \&
  Gharib]{Carmigniani2017}
{\sc \au{Carmigniani, R.~A.}, \au{Benoit, M.}, \au{Violeau, D.} \& \au{Gharib,
  M.}} \yr{2017}  \at{Resonance wave pumping with surface waves}.  \jt{J. Fluid
  Mech.}  \bvol{811},  \pg{1--36}.

\bibitem[Cho \& Kim(2000)]{Cho2000}
{\sc \au{Cho, I.~H.} \& \au{Kim, M.H.}} \yr{2000}  \at{Interactions of
  horizontal porous flexible membrane with waves}.  \jt{J. Waterw. Port Coast.
  Ocean Eng.}  \bvol{126}~(5),  \pg{245--253}.

\bibitem[Cho \& Kim(1998)]{Cho1998}
{\sc \au{Cho, I.~H.} \& \au{Kim, M.~H.}} \yr{1998}  \at{Interactions of a
  horizontal flexible membrane with oblique incident waves}.  \jt{J. Fluid
  Mech.}  \bvol{367},  \pg{139--161}.

\bibitem[Collins {\em et~al.\/}(2021)Collins, Hossain, Dettmer \&
  Masters]{Collins2021}
{\sc \au{Collins, I.}, \au{Hossain, M.}, \au{Dettmer, W.} \& \au{Masters, I.}}
  \yr{2021}  \at{Flexible membrane structures for wave energy harvesting: A
  review of the developments, materials and computational modelling
  approaches}.  \jt{Renew. Sustain. Energy Rev.}  \bvol{151},  \pg{111478}.

\bibitem[Desmars {\em et~al.\/}(2018)Desmars, Tchoufag, Younesian \&
  Alam]{Desmars2018}
{\sc \au{Desmars, N.}, \au{Tchoufag, J.}, \au{Younesian, D.} \& \au{Alam,
  M.-R.}} \yr{2018}  \at{Interaction of surface waves with an actuated
  submerged flexible plate: Optimization for wave energy extraction}.  \jt{J.
  Fluids Struct.}  \bvol{81},  \pg{673--692}.

\bibitem[Gayathri {\em et~al.\/}(2020)Gayathri, Benny \& Behera]{Gayathri2020}
{\sc \au{Gayathri, R.}, \au{Benny, C.} \& \au{Behera, H.}} \yr{2020} Wave
  attenuation by a submerged flexible permeable membrane.  \bt{In {\em AIP
  Conference Proceedings\/}}, ,  \vol{vol. 2277}. AIP Publishing.

\bibitem[Guo {\em et~al.\/}(2020)Guo, Mohapatra \& Soares]{Guo2020}
{\sc \au{Guo, Y.~C.}, \au{Mohapatra, S.~C.} \& \au{Soares, C.~G.}} \yr{2020}
  \at{Wave energy dissipation of a submerged horizontal flexible porous
  membrane under oblique wave interaction}.  \jt{Appl. Ocean Res.}  \bvol{94},
  \pg{101948}.

\bibitem[Kumar {\em et~al.\/}(2007)Kumar, Manam \& Sahoo]{Kumar2007}
{\sc \au{Kumar, P.~S.}, \au{Manam, S.~R.} \& \au{Sahoo, T.}} \yr{2007}
  \at{Wave scattering by flexible porous vertical membrane barrier in a
  two-layer fluid}.  \jt{J. Fluids Struct.}  \bvol{23}~(4),  \pg{633--647}.

\bibitem[Le~M{\'e}haut{\'e}(1976)]{LeMehaute1976}
{\sc \au{Le~M{\'e}haut{\'e}, B.}} \yr{1976} {\em An introduction to
  hydrodynamics and water waves\/}.  \publ{Springer Science \& Business Media}.

\bibitem[Lehmann {\em et~al.\/}(2013)Lehmann, Elandt, Pham, Ghorbani, Shakeri
  \& Alam]{Lehmann2013}
{\sc \au{Lehmann, M.}, \au{Elandt, R.}, \au{Pham, H.}, \au{Ghorbani, R.},
  \au{Shakeri, M.} \& \au{Alam, M.-R.}} \yr{2013} An artificial seabed carpet
  for multidirectional and broadband wave energy extraction: Theory and
  experiment.  \bt{In {\em Proceedings of 10th European Wave and Tidal Energy
  Conference\/}}.

\bibitem[Mirza {\em et~al.\/}(2024{\natexlab{{\em a\/}}})]{Mirza20241}
{\sc \au{Mirza, Tooba} \& \au{others}} \yr{2024{\natexlab{{\em a\/}}}}  \at{A
  novel approach for using submerged structure as wave-trapping zone}.
  \jt{Journal of Fluids and Structures}  \bvol{129},  \pg{104169}.

\bibitem[Mirza {\em et~al.\/}(2024{\natexlab{{\em b\/}}})]{Mirza20242}
{\sc \au{Mirza, Tooba} \& \au{others}} \yr{2024{\natexlab{{\em b\/}}}}  \at{A
  novel approach for using submerged structure as wave-trapping zone}.
  \jt{Journal of Fluids and Structures}  \bvol{129},  \pg{104169}.

\bibitem[Mohapatra \& Guedes~Soares(2019)]{Mohapatra2019}
{\sc \au{Mohapatra, S.~C.} \& \au{Guedes~Soares, C.}} \yr{2019}
  \at{Interaction of ocean waves with floating and submerged horizontal
  flexible structures in three-dimensions}.  \jt{Appl. Ocean Res.}  \bvol{83},
  \pg{136--154}.

\bibitem[Mohapatra \& Guedes~Soares(2020)]{Mohapatra2020}
{\sc \au{Mohapatra, S.~C.a} \& \au{Guedes~Soares, C.}} \yr{2020}
  \at{Hydroelastic response of a flexible submerged porous plate for wave
  energy absorption}.  \jt{J. Mar. Sci. Eng.}  \bvol{8}~(9),  \pg{698}.

\bibitem[Mohapatra \& Sahoo(2014)]{Mohapatra2014}
{\sc \au{Mohapatra, S.~C.} \& \au{Sahoo, T.}} \yr{2014}  \at{Wave interaction
  with a floating and submerged elastic plate system}.  \jt{J. Eng. Math.}
  \bvol{87}~(1),  \pg{47--71}.

\bibitem[Moisy {\em et~al.\/}(2009)Moisy, Rabaud \& Salsac]{Moisy2009}
{\sc \au{Moisy, F.}, \au{Rabaud, M.} \& \au{Salsac, K.}} \yr{2009}  \at{A
  synthetic schlieren method for the measurement of the topography of a liquid
  interface}.  \jt{Exp. Fluids}  \bvol{46}~(6),  \pg{1021--1036}.

\bibitem[Mork {\em et~al.\/}(2010)Mork, Barstow, Kabuth \& Pontes]{Mork2010}
{\sc \au{Mork, G.}, \au{Barstow, S.}, \au{Kabuth, A.} \& \au{Pontes, M.~T.}}
  \yr{2010} Assessing the global wave energy potential.  \bt{In {\em
  International Conference on Offshore Mechanics and Arctic Engineering\/}}, ,
  \vol{vol. 49118},  \pg{pp. 447--454}.

\bibitem[Ouellet \& Datta(1986)]{Ouellet1986}
{\sc \au{Ouellet, Y.} \& \au{Datta, I.}} \yr{1986}  \at{A survey of wave
  absorbers}.  \jt{J. Hydraul. Res.}  \bvol{24}~(4),  \pg{265--280}.

\bibitem[Pinon {\em et~al.\/}(2017)Pinon, Perret, Cao, Poupardin, Brossard \&
  Rivoalen]{Pinon2017}
{\sc \au{Pinon, G.}, \au{Perret, G.}, \au{Cao, L.}, \au{Poupardin, A.},
  \au{Brossard, J.} \& \au{Rivoalen, E.}} \yr{2017}  \at{Vortex kinematics
  around a submerged plate under water waves. part ii: Numerical computations}.
   \jt{Eur. J. Mech. B Fluids}  \bvol{65},  \pg{368--383}.

\bibitem[Poupardin {\em et~al.\/}(2012)Poupardin, Perret, Pinon, Bourneton,
  Rivoalen \& Brossard]{Poupardin2012}
{\sc \au{Poupardin, A.}, \au{Perret, G.}, \au{Pinon, G.}, \au{Bourneton, N.},
  \au{Rivoalen, E.} \& \au{Brossard, J.}} \yr{2012}  \at{Vortex kinematic
  around a submerged plate under water waves. part i: Experimental analysis}.
  \jt{Eur. J. Mech. B Fluids}  \bvol{34},  \pg{47--55}.

\bibitem[Renzi(2016)]{Renzi2016}
{\sc \au{Renzi, E.}} \yr{2016}  \at{Hydroelectromechanical modelling of a
  piezoelectric wave energy converter}.  \jt{Proc. R. Soc. A: Math. Phys. Eng.
  Sci.}  \bvol{472}~(2195),  \pg{20160715}.

\bibitem[Sheng(2019)]{Sheng2019}
{\sc \au{Sheng, W.}} \yr{2019}  \at{Wave energy conversion and hydrodynamics
  modelling technologies: A review}.  \jt{Renew. Sustain. Energy Rev.}
  \bvol{109},  \pg{482--498}.

\bibitem[Shoele(2023)]{Shoele2023}
{\sc \au{Shoele, K.}} \yr{2023}  \at{Hybrid wave/current energy harvesting with
  a flexible piezoelectric plate}.  \jt{J. Fluid Mech.}  \bvol{968},  \pg{A31}.

\bibitem[Stamos \& Hajj(2001)]{Stamos2001}
{\sc \au{Stamos, D.~G.} \& \au{Hajj, M.~R.}} \yr{2001}  \at{Reflection and
  transmission of waves over submerged breakwaters}.  \jt{J. Eng. Mech.}
  \bvol{127}~(2),  \pg{99--105}.

\bibitem[Stokes(1847)]{Stokes1847}
{\sc \au{Stokes, G.~G.}} \yr{1847}  \at{On the theory of oscillatory waves}.
  \jt{Trans. Cam. Philos. Soc.}  \bvol{8},  \pg{441--455}.

\bibitem[Ursell(1952)]{Ursell1952}
{\sc \au{Ursell, F.}} \yr{1952}  \at{Edge waves on a sloping beach}.  \jt{Proc.
  R. Soc. A: Math. Phys. Eng. Sci.}  \bvol{214}~(1116),  \pg{79--97}.

\bibitem[Wildeman(2018)]{Wildeman2018}
{\sc \au{Wildeman, S.}} \yr{2018}  \at{Real-time quantitative schlieren imaging
  by fast fourier demodulation of a checkered backdrop}.  \jt{Exp. Fluids}
  \bvol{59}~(6),  \pg{1--13}.

\end{thebibliography}

	\end{document}